# Quantifying Twist Angles in Cuprate Heterostructures with Anisotropic Raman Signatures

Flavia Lo Sardo,[1, 2] Marina Esposito,[3, 4] Tommaso Confalone,[1, 5] Christophe Tremblay,[6]
Valerii M. Vinokur,[7] Genda Gu,[8] Domenico Montemurro,[3] Davide Massarotti,[9]
Francesco Tafuri,[3] Kornelius Nielsch,[1, 5, 2] Nicola Poccia,[1, 3, *] and Golam Haider[1, †]

[1]Leibniz Institute for Solid State and Materials Research Dresden (IFW Dresden), 01069 Dresden, Germany
[2]Institute of Materials Science, Technische Universitaet Dresden, 01062 Dresden, Germany
[3]Department of Physics, University of Naples Federico II, 80126 Naples, Italy
[4]National Institute for Nuclear Physics (INFN) - Sezione di Napoli, 80126, Naples, Italy
[5]Institute of Applied Physics, Technische Universit¨at Dresden, 01062 Dresden, Germany
[6]Department of Physics, Universit´e de Sherbrooke, J1K2R1 Sherbrooke (Qu´ebec), Canada
[7]Terra Quantum AG, 9000 St. Gallen, Switzerland
[8]Condensed Matter Physics and Materials Science Department,
Brookhaven National Laboratory, Upton, NY 11973, USA
[9]Department of Electrical Engineering and Information Technology,
University of Naples Federico II, 80125 Naples, Italy

Artificially engineered twisted van derWaals (vdW) heterostructures have unlocked new path-
ways for exploring emergent quantum phenomena and strongly correlated electronic states. Many
of these phenomena are highly sensitive to the twist angle, which can be deliberately tuned to tailor
the interlayer interactions. This makes the twist angle a critical tunable parameter, emphasizing
the need for precise control and accurate characterization during device fabrication. In particular,
twisted cuprate heterostructures based on $Bi_2Sr_2CaCu_2O_{8+x}$ have demonstrated angle-dependent
superconducting properties, positioning the twist angle as a key tunable parameter. However, the
twisted interface is highly unstable under ambient conditions and vulnerable to damage from con-
ventional characterization tools such as electron microscopy or scanning probe techniques. In this
work, we introduce a fully non-invasive, polarization-resolved Raman spectroscopy approach for de-
termining twist angles in artificially stacked BSCCO heterostructures. By analyzing twist-dependent
anisotropic vibrational Raman modes, particularly utilizing the out-of-plane $A_{1g}$ vibrational mode
of Bi/Sr at $\sim 116 \ cm^{-1}$, we identify clear optical fingerprints of the rotational misalignment between
cuprate layers. Our high-resolution confocal Raman setup, equipped with polarization control and
RayShield filtering down to $10 \ cm^{-1}$, allows for reliable and reproducible measurements without
compromising the material's structural integrity.

## I. INTRODUCTION

Two-dimensional (2D) materials have emerged as a
cornerstone of modern condensed matter physics and
quantum materials research. Since the isolation of
graphene in the early 2000s [1], the materials science
community has witnessed an explosion in the discovery
and synthesis of 2D crystals including semiconductors [2],
magnets [3], topological insulators [4], and superconduc-
tors [5]. The key to their versatility lies in the nature
of their weak interlayer coupling. Thanks to the van
der Waals (vdW) interactions atomically thin sheets can
be exfoliated and restacked without the need for lattice
matching [6, 7], enabling the creation of designed het-
erostructures with novel properties[8, 9]. The advent of
these artificially stacked 2D layers with controlled ori-
entation and interface chemistry has opened a new fron-
tier in quantum material design [10, 11]: these structures
can exhibit emergent properties stemming from interlayer
coupling [12], quantum confinement [13], moir´e physics

[14], or symmetry breaking [15] which are unattainable
in bulk crystals [9]. By controlling the twist angle be-
tween layers, for instance, it is possible to engineer flat
bands [16], topological states [17], and unconventional
superconductivity [18, 19].

One of the most important bulding blocks of these
heterostructures are atomically thin superconductors,
such as NbSe₂ [20], TaS₂ [21] and $Bi_2Sr_2CaCu_2O_{8+x}$
(BSCCO) [22]. Among these, the vdW layered high-
$T_c$ cuprate, BSCCO holds particular promise. Unlike
conventional superconductors, BSCCO retains its intrin-
sic superconductivity even when exfoliated down to a
few unit cells [23]. Moreover, the presence of d-wave
pairing symmetry enables unprecedented control over su-
perconductivity in twisted BSCCO Josephson junctions
[18, 19, 24]. At the same time, BSCCO remains a chal-
lenging material due to its high sensitivity to thermal
degradation and chemical exposure, especially in thin-
film form, often resulting in a complete loss of super-
conducting properties [25]. This highlights that not only
the fabrication process and the control of the twist an-
gle, but also the accurate determination of the twist-
ing angle present significant technical challenges [25, 26].
Unlike in conventional structures made of graphene or transition metal dichalcogenide monolayers, where twist angles can be confirmed using electron diffraction [27], second-harmonic generation (SHG) [28], or atomic force microscopy (AFM) [29, 30], such approaches are often impractical for twisted BSCCO heterostructures due to their environmental sensitivity issue. Moreover, the optically complex nature of BSCCO layers, along with the absence of intrinsic inversion symmetry breaking, limits the applicability of established twist-angle characterization methods, such as SHG. These factors present a substantial obstacle to accurately determining the twist angle in artificially assembled heterostructures of ultrathin, fragile layers.

* nicola.poccia@unina.it
† g.haider@ifw-dresden.de

An alternative, non-destructive method is polarized Raman spectroscopy. Polarization-dependent measurements have long been employed as a sensitive probe of symmetry-breaking phenomena across a wide range of material platforms, including copper oxide superconductors, where Raman selection rules have been used to detect stripe ordering and electronic anisotropies [31], as well as in van der Waals materials such as $MoS_2$ [32] and hexagonal boron nitride [33], where polarization resolved Raman scattering reveals symmetry breaking and anisotropic responses. Not only, this technique has been successfully applied to extract local crystal orientation based on the anisotropy of Raman mode intensities and to demonstrate optical anisotropy in crystalline materials, such as silicon, black phosphorus and CrSBr [34–37]. In high-$T_c$ cuprates, local anisotropy has been attributed to the combined effects of oxygen ordering, electronic nematicity, and subtle lattice distortions [38]. Single-crystal X-ray diffraction measurements on BSCCO indicate a pseudo-tetragonal unit cell with lattice parameters that reflect a slight orthorhombicity [39]. In particular, high-resolution STM measurements revealed a subtle orthorhombic distortion in the Bi–O surface layer, where two Bi sublattices are displaced by approximately 1% of the unit cell along the orthorhombic a-axis [40]. This distortion breaks inversion symmetry at the Cu sites while preserving a mirror plane, although the overall crystal retains three-dimensional inversion symmetry, with the inversion center located within the calcium layer. Complementary synchrotron X-ray diffraction studies [41] identified orthorhombic structural domains embedded within the nominally tetragonal matrix of BSCCO. These domains, marked by in-plane distortions and broken rotational symmetry, provide compelling structural evidence for the presence of local anisotropy consistent with electronic nematicity.

Building on this deviation from ideal tetragonal symmetry, we utilize the in-plane anisotropic modulation of Raman-active phonon intensities under linearly polarized excitation to determine the crystalline orientation of BSCCO. By mapping the angular dependence of Raman mode intensities, we are able to clearly identify the relative orientation of Cu–O bond axes that enables to quantify twist angle between flakes and resolve local symmetry breaking in BSCCO layers and heterostructures. This es-tablishes polarization-resolved Raman spectroscopy as a powerful, non-destructive method for twist-angle determination in BSCCO-based heterostructures, while also providing insight into subtle crystalline anisotropies.

## II.  RESULTS AND DISCUSSION

**Figure 1A** shows an optical image of a BSCCO flake exfoliated onto a Si/SiO$_2$. The inset displays a schematic of the BSCCO crystallographic unit cell, highlighting the layered structure of this compound. The unit cell extends ∼3.07 nm along the c-axis and reveals an alternation of superconducting and insulating layers, making BSCCO an intrinsic Josephson junction material. The weak vdW interaction between adjacent BiO planes facilitates mechanical exfoliation using standard scotch tape-based methods.

Following exfoliation, the structural properties of the BSCCO flakes were investigated through Raman spectroscopy using a customized polarization-resolved confocal setup, as illustrated in Figure 1B. A linearly polarized 532 nm excitation source of excitation power 3.5 mW was used to obtain the Raman spectra as shown in Figure 1C. Using a 532 nm polarization rotator, the laser polarizer was aligned to maximize the visibility of the principal Raman-active modes of BSCCO. A Bragg notch filter in the detection path enabled simultaneous collection of both Stokes and anti-Stokes signals, extending down to ±10 cm$^{-1}$ from the Rayleigh scattering line.

The recorded spectrum in Figure 1C reveals several characteristic phonon features of BSCCO, alongside minor contributions from the underlying silicon substrate. Mode assignments were made with reference to prior Raman spectra studies of BSCCO reported in the literature [42–49]. In a nutshell, in BSCCO, the sharp Raman peaks observed across the spectrum reflect a range of lattice vibrations involving both heavy atoms and oxygen-related modes. At low frequencies, a distinct peak around 27 cm$^{-1}$ is attributed to the acoustic mode, i.e., the mode induced by the superstructural modulation [43]. Two intense peaks at approximately 50 cm$^{-1}$ and 60 cm$^{-1}$ are commonly associated with in-plane vibrations involving heavy Bi atoms, as reported in multiple studies [42–48]. A prominent mode centered at ∼ 116 cm$^{-1}$, with visible shoulders at ∼ 106 cm$^{-1}$ and ∼128 cm$^{-1}$, is attributed to out-of-plane vibrational modes involving Bi and/or Sr atoms with respect to the CuO$_2$ plane [43, 44, 50]. The peak near 190 cm$^{-1}$ has been linked to either Sr or Cu vibrations [44, 45, 49]. In the region between 290 and 300 cm$^{-1}$, other peaks are visible which can be attributed to a higher-frequency BSCCO mode referring to vibrations of O bonded to Cu or Bi [46, 47]. In the higher-frequency region, several broad and complex modes arise, primarily related to oxygen dynamics. The ∼ 290–300 cm$^{-1}$ mode corresponds to an A$_{1g}$ phonon involving c-axis (out-of-plane) vibrations of in-plane oxygen atoms within the CuO$_2$ layers [46, 47], while the ∼

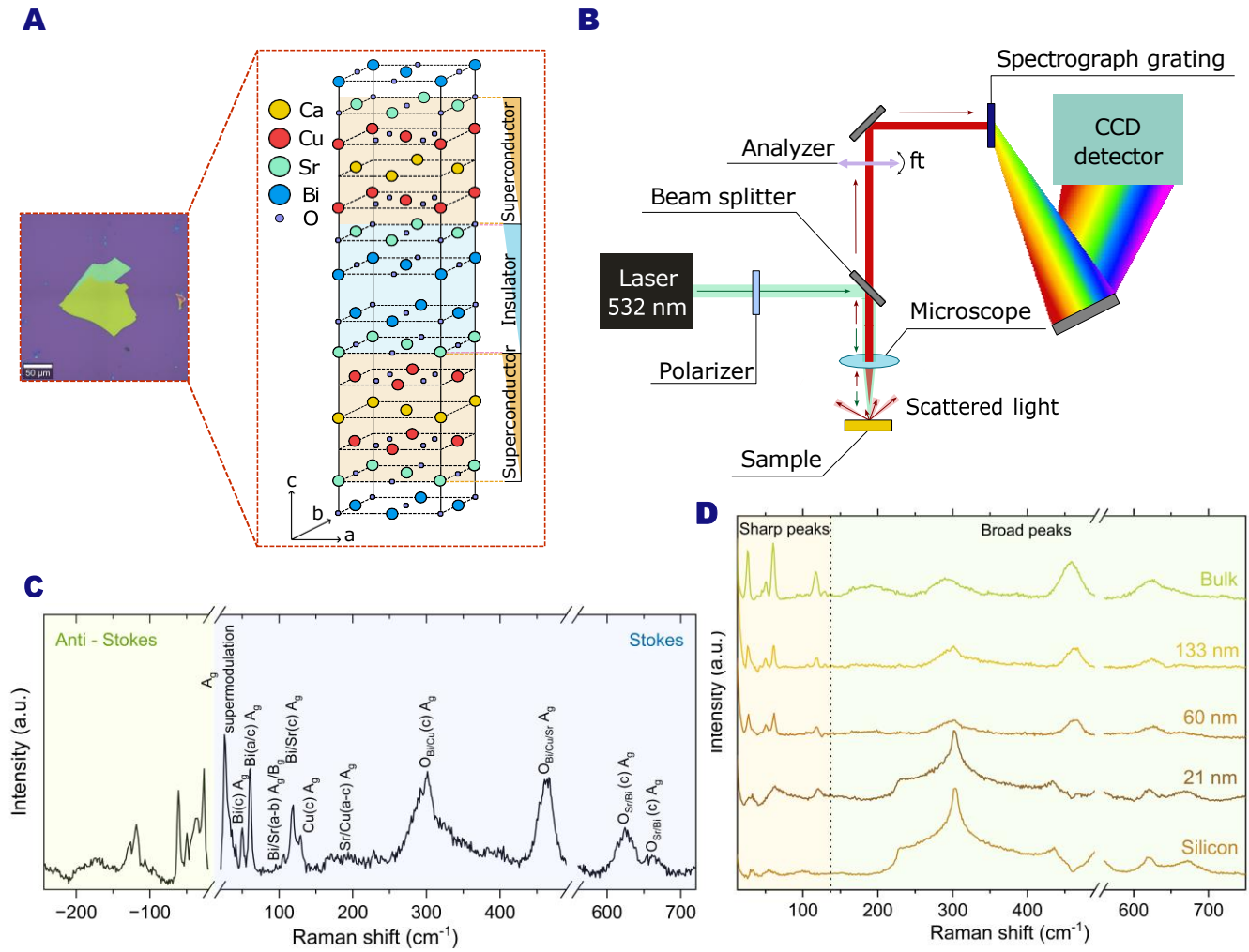

FIG. 1. Micro Raman spectroscopy on BSCCO flakes of different thicknesses. (A) Optical image of a BSCCO flake exfoliated on a Si/SiO$_2$ substrate. The inset is a schematic representation of BSCCO unit cell. (B) Schematic of the micro Raman spectrometer setup used in this study, featuring a 532 nm laser, two Bragg notch filters incorporated beam splitter, and the spectrometer containing a depolarizer. The polarization of the incident and scattered light is independently controlled using a polarizer and an analyzer, respectively. (C) Raman spectrum acquired on BSCCO in HH (parallel polarization of the incident and scattered light) configuration at 3.5 mW laser power. The spectrum spans the range -240 cm$^{-1}$ to +720 cm$^{-1}$ including both anti-Stokes and Stokes regions. (D) Raman spectra of BSCCO flakes with thicknesses ranging from 22 nm to bulk and Si/SiO$_2$ substrate taken in HH configuration at 3.5 mW. Modes up to 116 cm$^{-1}$ exhibit sharp and well-defined features, while higher-frequency modes appear broader and less resolved.

460 cm$^{-1}$ mode originates from an A$_{1g}$ phonon involving c-axis vibrations of oxygen atoms, especially those in the Bi–O layers [43, 51, 52]. Finally, the broad modes observed near $\sim$ 625 cm$^{-1}$ and $\sim$ 660 cm$^{-1}$ are attributed to in-plane bond-stretching vibrations of oxygen atoms in the Sr–O layers, and are often enhanced or activated by structural modulation and oxygen non-stoichiometry [43, 51, 52].

The evolution of these peaks is then studied as a function of the flake thickness. The spectra of flakes with a thickness ranging from bulk ($\sim$ 370nm) to 22 nm were collected and compared. To have a reference for the thin-ner flake the Raman spectrum of the Si/SiO$_2$ substrate was also recorded. The Spectra are shown in Figure 1D. The flake thicknesses were determined by atomic force microscopy (AFM). The result are presented in **Figure S1, Supporting Information**. All spectra were acquired at 3.5 mW in parallel configuration, and the spectral window restricted to the Stokes scattering region to enhance the visibility of key Raman features. Flakes from bulk down to 60 nm exhibit the full set of characteristic Raman peaks of BSCCO. When reaching 22 nm, however, noticeable changes in the vibrational modes emerge, as also reported in other works [53]. In this case, the

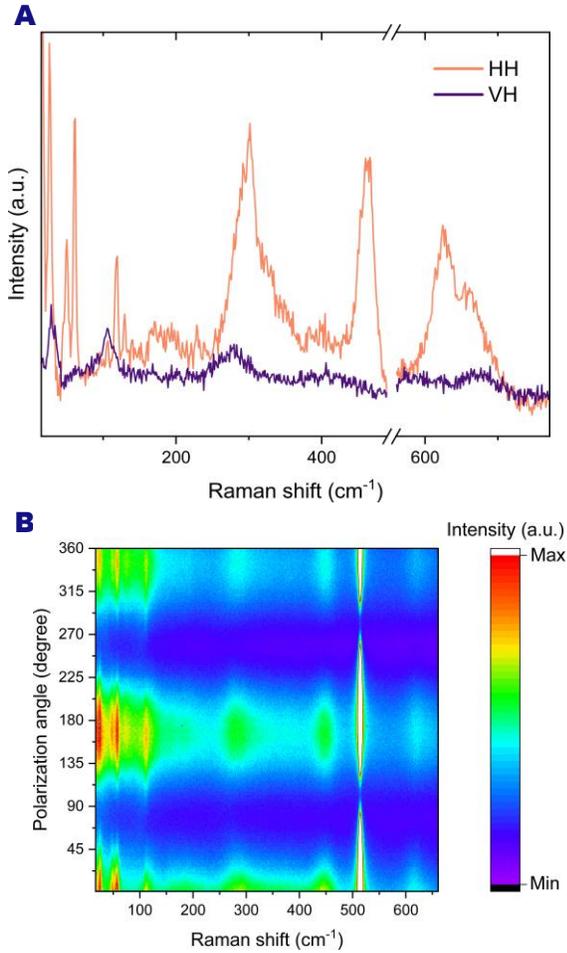

**A**

**B**

FIG. 2. Polarization resolved Raman spectroscopy on a BSCCO flake randomly oriented with respect to the polarization of the excitation field. (A) Parallel (HH) and cross (VH) polarized Raman spectra of a $\sim$ 60 nm BSCCO flake. (B) Color map of the Raman mode intensity in parallel (HH) configuration while the incident laser polarizer and analyzer were continuously rotated across 0° to 360°.

Raman mode intensity of the substrate becomes dominant, making the spectral features of BSCCO less distinguishable. Additionally, as thickness decreases, the layers lose long-range interlayer coherence, which is essential for sustaining collective vibrational modes—particularly the low-frequency acoustic phonons associated with supermodulation and those characterized by out-of-plane motion. The $\sim$ 460 cm$^{-1}$ mode, involving oxygen vibrations perpendicular to the CuO$_2$ plane, depends on restoring forces from adjacent layers (SrO or Ca). Thinning down reduces these forces, weakening the vibrational coherence and Raman activity. Similarly, the modes at $\sim$ 50 cm$^{-1}$ and $\sim$ 60 cm$^{-1}$ merge into a single broader feature. On the contrary, modes in which the silicon substrate contributes, such as those near $\sim$ 300 cm$^{-1}$ and $\sim$ 660 cm$^{-1}$ become more pronounced due to the reduced thickness of the BSCCO layer. However, in agreement with pre-vious reports, [53], the peak at $\sim$ 116 cm$^{-1}$ exhibits a shift as the flake becomes thinner, reaching approximately 120 cm$^{-1}$ for the 22 nm-thick sample. To further verify the stability of this peak across a given flake, a spatial mapping was performed (**Figure S2**), confirming its consistency across different locations. This behavior is attributed to mechanical strain induced by substrate interactions and the reduced bulk support in thinner flakes [53]. Importantly, this peak persists across all measured thicknesses, regardless of the sample history and aging, making it a stable signature for identifying BSCCO even in very thin flakes. The polarization-resolved Raman spectra in **Figure 2A** exhibit strong polarization selectivity of the Raman modes. Before acquiring the data, the polarization of the incident light was adjusted, ensuring maximum intensity of the representative Raman modes. This is due to the single-crystalline nature of the sample. Then, rotating the polarizer by 90°, parallel and cross-polarized Raman spectra were recorded. Interestingly, most of the Raman modes become significantly suppressed or even disappear entirely, exhibiting a very high degree of polarization. This strong anisotropy of the Raman modes, is consistent with literature [44, 46].

To investigate this polarization dependence of the anisotropic modes in greater detail, a polarization-resolved measurement in the HH configuration was performed by rotating the polarizer from 0° to 360°. The color plot of the obtained Raman intensity as a function of polarization angle is shown in Figure 2B. The color scale on the right side represents the magnitude of the Raman mode intensity. This representation highlights the evolution of mode intensities with respect to the incident polarization direction. The sharp peaks at low frequencies ($\sim$ 50 cm$^{-1}$, $\sim$ 60 cm$^{-1}$, $\sim$ 116 cm$^{-1}$) exhibit periodic sinusoidal modulation. The observed anisotropy of the $\sim$ 116 cm$^{-1}$ Raman mode in backscattering geometry arises from subtle orthorhombic modulation that breaks the in-plane rotational symmetry of BSCCO. Although the crystal is nominally pseudo-tetragonal, high-resolution structural studies (e.g. XRD, STM, synchrotron) have revealed orthorhombic distortions, particularly in the Bi–O surface layer, where Bi atoms are displaced along the a-axis [38, 40, 41, 54]. This distortion breaks the fourfold ($C_4$) rotational symmetry, leading to the polarization-dependent Raman intensity of the 116 cm$^{-1}$ A$_{1g}$ mode, an effect that would be absent in a perfectly isotropic ab-plane. This behavior can be explained by assuming complex values for the Raman tensor elements and taking into account the orthorhombic symmetry of the crystal structure [55]. Similar periodic behavior is also observable for the other low-frequency sharp modes, such as those around 50-60 cm$^{-1}$. However, the thickness-dependent Raman spectra reveal that several external factors, including strain, perturb these modes. Regarding Cu-O related modes in the 300–600 cm$^{-1}$ range, although showing periodic modulation of intensity, they are broad and often consist of overlapping features. Such overlap may carry different angular depen-

**A**

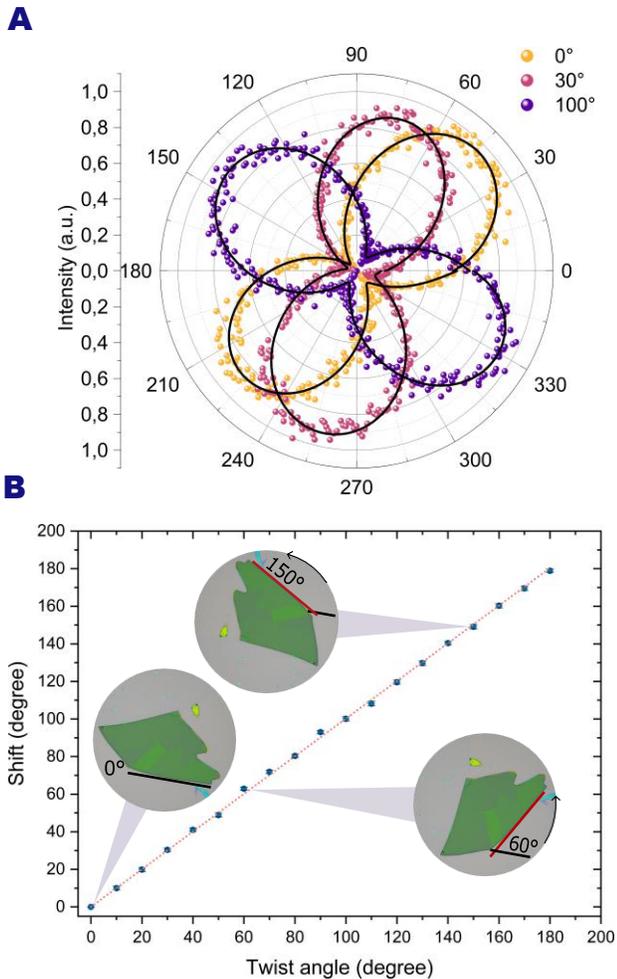

**B**

FIG. 3. Polarization resolved Raman spectra for finding crystalline orientation. (A) Polar plots of the angular dependence of 116 cm⁻¹ mode for three different rotation angles (0°, 30°, 100°). The phase shift of the lobes directly reflects the mechanical rotation applied via the goniometer. (B) Extracted angular shift of the Raman pattern as a function of the applied twisting angle. The experimental data (blue dots) follow a linear trend (red fit), confirming the ability of Raman spectroscopy to track the in-plane orientation of the flake. Insets show optical images of the sample at three different orientations (0°, 60°, 150°).

dences, which can induce larger deviations in twist-angle determination. In light of these observations, and the already mentioned robustness with respect to thinning, sample aging, as well as the polarization dependence and sharpness, the ∼ 116 cm⁻¹ mode is suitable for investigating crystalline orientation by spectroscopic Raman measurement technique.

We now turn to determining the relative crystalline orientation of BSCCO flakes using polarization-resolved Raman spectroscopy. A Si/SiO₂ substrate with exfoliated BSCCO flakes on top of it was mounted on a goniometer

(**Figure S3**) integrated with a motorized Raman stage beneath the objective lens. The excitation laser polarization was fixed along the x-axis of the stage, and the sample was rotated around the laser beam axis. This configuration enabled efficient angular scans while ensuring consistent and reproducible measurements.

A bulk-like flake (∼ 300 nm) with an extended surface was selected to allow multiple acquisitions on fresh regions, minimizing potential laser-induced damage. The laser power was set to 6 mW, which provided high signal quality without causing detectable degradation. To further enhance the signal-to-noise ratio, the polarizer and analyzer were decoupled, allowing greater light throughput and reduced background noise. To obtain the relative crystalline orientation of a flake due to the twist induced by the goniometer, Raman measurements were performed at 18 different twist angles, ranging from 0° to 180°. For each twist angle, a set of 360 Raman spectra was acquired by rotating the excitation polarization from 0 to 360 degrees. For each twist angle, the periodic modulation of the ∼ 116 cm⁻¹ mode was extracted from the spectrum and fitted with a sinusoidal function (**Figure S4**).

**Figure 3A** shows a polar plot illustrating the intensity profile of the ∼ 116 cm⁻¹ mode at three representative twist angles. The corresponding sinusoidal fits are shown in **Figure S5**. The original position of the flake with respect to the laser polarization was assigned to 0° twist angle, as shown in orange color. Interestingly, the polar plots of the other two profiles, while the sample was twisted by 30° (purple) and 100° (orange), reflect a shift of the obtained maxima around 30° and 100°, i.e., the twist angles of the flake. Hence, the obtained shift of the maxima of the polar/sinusoidal plots of randomly oriented 2D BSCCO crystals can provide the relative orientation between the flakes.

We consolidate this argument by performing such measurements at 18 different twist angles. The twist angle was then determined by evaluating the difference in phase (i.e., the position of the maxima or minima) between each fitted curve and the reference. Figure 3B shows a plot of the extracted shift versus the real twist angle. The data reveal a clear linear correlation between the angle obtained by the Raman measurement technique and the twist angle obtained from the goniometer. The optical microscopy images of the flake used for measurements under different twist angles are provided in the insets.

Leveraging the capabilities of polarization-resolved Raman spectroscopy, we determined the twist angle in two artificially engineered twisted BSCCO heterostructures. The illustrated heterostructures in **Figure 4A** and **4C** were fabricated using our in-house cryogenic exfoliation and stacking procedure [19, 24], which ensures minimal contamination and excellent interfacial contact between the flakes, while protecting the twisted interface preserved from thermal degradation. Detailed information about the fabrication protocol can be found in the Meth-

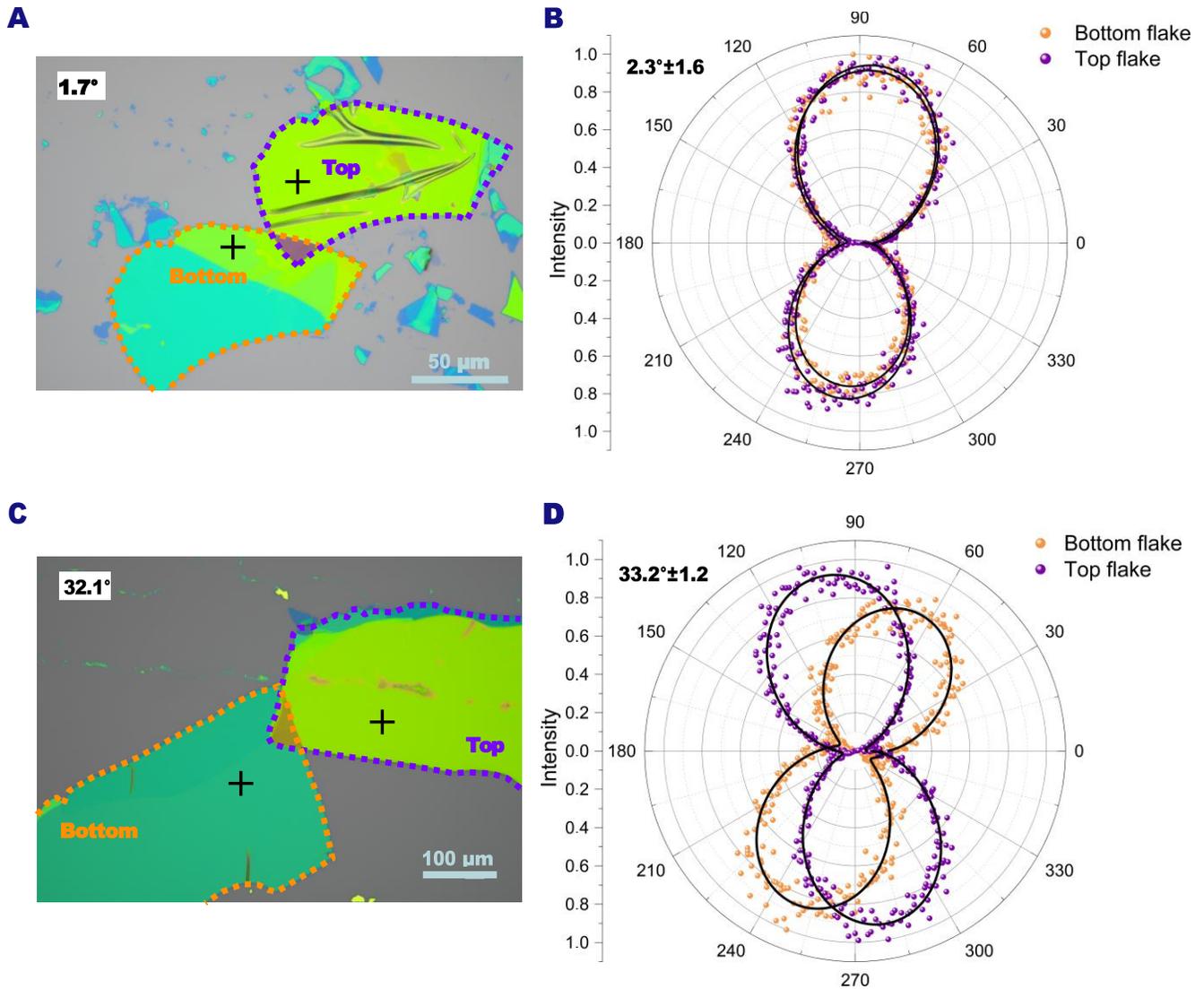

FIG. 4. Raman − based detection of twist angle in BSCCO−BSCCO vdW structures. (A, C) Optical microscope images of two BSCCO−BSCCO heterostructures fabricated with twist angles of approximately 3° and 33°, respectively. The individual flakes forming each junction are outlined with dotted lines: orange for the bottom flake and purple for the top flake. The positions of Raman measurements are marked with a cross. (B, D) Polar plots of the Raman intensity of the $\sim 116$ cm$^{-1}$ A$_{1g}$ mode as a function of the polarization angle, acquired independently on each flake in (A) and (C). The angular position of the intensity maxima reveals a relative shift between the two plots, corresponding to the twist angle between the flakes.

ods section. To determine the twist angle between the top and bottom BSCCO flakes, we recorded polarization-resolved Raman spectra by continuously rotating the polarization direction of the excitation laser from 0° to 360°. For each junction, a corresponding polar plot is presented (Figure 4B and 4D), exhibiting an angular shift of approximately 2° (2.3 ±1.6) and 33° (33.2 ±1.2), respectively. These angular shifts match well with the estimated twist angles obtained from optical microscopy (i.e. 1.7° and 32.1°), validating the reliability of the Raman-based approach. The distinct angular dependence of the

Raman intensity originates from the apparent twist angle between the layers, which modifies the polarization selection rules. The slight difference in overall integrated intensity visible in the ~33° device is attributed to the different thicknesses of the flakes (i.e., ~40 nm bottom flake, ~105 nm top flake ), as evidenced by their two distinct colors. This is not the case for the ~3° device, as the two flakes have similar thicknesses (i.e., both flakes are ~90 nm).

In the heterostructures examined here, both the top and bottom flakes originate from the same parent crys-tal. This allows the twist angle to be directly inferred from their relative orientation in the optical image. However, in more complex heterostructures assembled from flakes exfoliated from different parent crystals, this optical imaging approach becomes ineffective due to the lack of a common crystallographic reference. In such scenarios, the polarization-resolved Raman method remains applicable and indispensable.

It is also important to highlight that BSCCO maintains inversion symmetry in its bulk form, rendering conventional nonlinear optical techniques such as second harmonic generation (SHG) unsuitable for probing twist angles in this material. In contrast, the polarization dependence of Raman-active phonon modes provides a sensitive and non-destructive alternative to extract the twist angle, independent of the inversion symmetry constraints.

## III. OUTLOOK

The demonstrated polarization-resolved Raman technique provides a robust pathway for reliable twist-angle determination in van der Waals heterostructures of similar and dissimilar materials, particularly in centrosymmetric systems such as BSCCO, where conventional structural characterization methods often fail. Beyond this specific case, the approach can be extended to a broad class of layered quantum materials, including cuprates, ruthenates, and iron-based superconductors. An important next step would be the implementation of spatially resolved Raman mapping, which could uncover local twist-angle variations arising from crystalline inhomogeneity, interfacial impurities, or lattice-mismatch-induced strain [30], where microscopic twist-angle domains can strongly influence correlated electronic states. Furthermore, integrating this technique with in situ cryogenic transport setups would allow direct exploration of twist-angle-dependent superconducting, charge-density-wave, or pseudogap phenomena under well-controlled conditions. Overall, these developments would not only deepen our understanding of twist-angle tuned correlated physics but also provide a versatile platform for engineering novel quantum devices in which rotational degrees of freedom serve as active and precise tuning parameters.

## IV. CONCLUSION

We have demonstrated that Raman spectroscopy offers a rapid, non-destructive, and reliable technique for determining the twist angle in BSCCO-based heterostructures. This approach leverages the pronounced anisotropic angular dependence of Raman-active phonon modes, particularly the $\sim 116$ cm$^{-1}$ mode in BSCCO, which exhibits a robust and periodic intensity modulation that remains consistent across varying flake thicknesses and aging conditions. The method was first validated using twisted BSCCO flakes and then successfully applied to complex BSCCO-based twisted structures. Importantly, this technique is not limited to homostructures but it is generalizable to twisted heterostructures comprising dissimilar materials, where twist angle determination is challenging or inaccessible only via fabrication alignment. Therefore, Raman-based twist angle metrology provides a powerful and versatile tool for the characterization of environmentally sensitive quantum materials and devices, paving the way for more controlled studies of twist-dependent phenomena in low-dimensional superconductors and correlated systems.

## V. METHODS SECTION

*Substrate preparation:* Si/SiO$_2$ substrates with a 285 nm oxide layer were first cleaned sequentially with acetone and isopropanol to remove surface impurities. Subsequently, they were placed in a glovebox environment for storage. To further enhance adhesion and eliminate organic contaminants, the substrates were treated with O$_2$ plasma prior to flake transfer. After cleaning, the substrates were baked at 150 °C for two hours to remove residual moisture.

*Exfoliation:* BSCCO bulk flakes (T$_c$ $\sim$91 K, $\Delta$T$_c$ $\sim$2 K [56, 57]) flakes were mechanically exfoliated from optimally doped single crystals using the standard Scotch tape method. To minimize polymer residues, the exfoliated flakes were transferred onto the prepared substrates using SPV 5057A5 Nitto tape.

*Twisted heterostructures fabrication:* Suitable BSCCO flakes were first identified by optical contrast, and the sample stack was cooled to -90°C. A poly(dimethylsiloxane) (PDMS) stamp mounted on a glass slide and positioned with a micromanipulator was brought into contact with the edge of the flake, after which the system was allowed to thermalize. Rapid detachment of the stamp cleaved the crystal along a flat plane between BiO layers, producing two thinner flakes. The flake remaining on the PDMS stamp was aligned and placed back onto the bottom flake on the substrate. The stack was then gradually warmed to -30°C. Further details on the device fabrication procedure can be found in Refs.[19, 24].

*Raman spectroscopy:* Raman spectroscopy was performed in a backscattering geometry using a WITec Alpha 300 confocal Raman microscope equipped with a 100× objective lens of NA = 0.9 at room temperature. A continuous-wave 532 nm laser was employed as the excitation source. Polarization-resolved measurements were conducted by independently varying the polarization states of both the excitation and the detected scattered light.

To achieve high spectral resolution, a diffraction grating with 1800 lines/mm was utilized, yielding a spectral resolution of approximately 1.2 cm$^{-1}$. A 532 nm RayShield filter was employed to enable the detection of low-frequency Raman modes down to 10 cm$^{-1}$ on both the Stokes and anti-Stokes sides of the spectrum.

To ensure accurate polarization analysis, a depolarization filter was placed in the detection path prior to the monochromator. This setup allowed for the reliable measurement of polarization-dependent Raman scattering signals with minimized influence from spurious polarization components. The laser power at sample was measured with a calibrated power meter maintained at 3.5 mW for the single spectra and thickness dependent measurements, and increased to 6 mW for the flakes in twisted heterostructures in order to improve the signal-to-noise ratio. The latter level was verified not to cause any relevant damage or degradation of the BSCCO flake over an acquisition time of 3 s per spectrum, repeated over 361 points (i.e., from 0° to 360°). Angular-dependent measurements were performed by rotating the sample in steps of 10° by using the rotational goniometer stage (angular resolution of 1°).

*Atomic force microscopy:* A Dimension Icon AFM system (Bruker) operating in tapping mode, equipped with TESPA-V2 cantilevers, was used to determine the thickness of the BSCCO flakes. Thickness values were estimated from single-point height profiles acquired on each flake using Gwyddion software.

*Data analysis:* Data were analyzed using Origin software for both the representation of Raman spectra and the evaluation of periodic trends and angular shifts. For the shifts calculation, the raw data were fitted with a sinusoidal function of the form:

$$y = y_0 + A \sin\left(\pi \frac{(x - x_c)}{w}\right) \tag{1}$$

where $y_0$ is the baseline, $A$ the amplitude, $x_c$ the phase shift, and $w$ half the oscillation period. The maxima were calculated from the fit parameters as:

$$x_{max} = x_c + \frac{w}{2} \tag{2}$$

The associated uncertainty was obtained by error propagation of the fit parameters:

$$\sigma_{x_{max}} = \sqrt{\sigma_{x_c} + \left(\frac{\sigma_w}{2}\right)^2} \tag{3}$$

*Error analysis:*

In addition to the uncertainty estimated from the fit parameters, potential sources of error include instrumental contributions such as the finite spectral resolution of the Raman spectrometer, the resolution of the polarizer and analyzer elements, and the angular precision of the goniometer integrated into the Raman microscope stage. Sample heterogeneity was evaluated and ruled out as a significant factor, since measurements at different positions across the flakes yielded highly consistent results (see **Figure S6**). Possible systematic effects, such as laser power fluctuations or residual background signals, were further minimized through repeated measurements and careful calibration. Overall, the reported uncertainty primarily reflects the limits of the fitting procedure in combination with these instrumental factors, while sample-related or extrinsic effects are negligible.

**Acknowledgements.** N.P. acknowledges the partial funding by the European Union (ERC-CoG, 3DCuT, 101124606), by the Deutsche Forschungsgemeinschaft (DFG 512734967, DFG 492704387, DFG9539383397, DFG 460444718, and DFG 452128813). N.P. and V.M.V. acknowledge support from Terra Quantum AG. Work at Brookhaven national laboratory is supported by US DOE (DE-SC0012704). F.T. and D. M. would like to acknowledge the following project: PNRR MUR project PE0000023 NQSTI. F.L.S. and G.H. thank Matthias Finger for his valuable assistance with the Raman setup, and Christiane Kranz for technical support in the laboratory.
**Author contributions.** N.P. and G.H. conceived and designed the experiment. F. L. S performed the experiments and analyzed the data with the contribution of M.E., T. C., C. T. The cuprate crystals were provided by G. G. The fabrication procedures and measurements were discussed by F. L. S., T.C., G.H., N. P. and F. L.S., M. E., T. C., V. M. V., D. M., D. M., F. T., K. N., G. H, N. P discussed the results. F. L. S., N. P., G. H. wrote the manuscript.
**Declaration of Interest** All authors declare no conflict of interest.